\documentclass[11pt]{article}
\usepackage[margin=0.82in]{geometry}
\usepackage{microtype}
\usepackage{amsmath,amssymb}
\usepackage{booktabs,tabularx,array,multirow}
\usepackage{graphicx}
\usepackage{caption,subcaption}
\usepackage{tikz}
\usetikzlibrary{arrows.meta,positioning,fit,calc,shapes.geometric}
\usepackage{pgfplots}
\pgfplotsset{compat=1.18}
\usepackage[numbers,sort&compress]{natbib}
\usepackage[hidelinks]{hyperref}
\usepackage[nameinlink,noabbrev]{cleveref}
\usepackage{xcolor}
\usepackage{enumitem}
\setlist{nosep,leftmargin=*}
\title{\textbf{Maat: Independent Deterministic Contract-Based Governance for Multi-Agent LLM Workflows}}
\author{Uliana Elina\\SynWe Group s.r.o.\\\texttt{uliana.elina@synwe.ai}}
\date{September 2026}

\begin{document}
\maketitle

\begin{abstract}
Large-language-model multi-agent systems (LLM-MAS) introduce a characteristic reliability problem: an error produced by one agent can be accepted as context by downstream agents and propagate across the workflow. Recent studies have documented inter-agent misalignment, error cascades, silent trajectory failures, and architecture-dependent error amplification, motivating verification at agent boundaries rather than only on final outputs. Many proposed safeguards, however, rely on learned or LLM-based judges whose verdicts are themselves probabilistic and can inherit the failure modes they are meant to catch; we ask whether a deterministic layer can instead stop contract-detectable handoff defects. We present \textbf{\emph{Maat}}, a runtime governance layer that validates agent-to-agent handoffs against a versioned workflow contract, or \emph{anchor}, with no language model in the validation or scoring path. We evaluated the approach in six controlled domain workflows (6--15 agents, 522 trials) with injected data-level defects and a deterministic seven-check rubric. Version~1 of this article reported rubric gains in all six workflows (2.9--26.5\%). A post-publication audit found that three benchmark scorers credited any early halt as a prevented defect, so false-alarm halts were counted as catches. On paired trials in which the governed run either completed or halted on a finding attributable to a verified defect, the rubric score changes by +7.7\% to +29.1\% in five workflows and is flat in software development (+0.8\%); model-call cost falls by 17--53\% where attributable halts occur early. A hand review of all 94 governed-arm halts found 35 false alarms (37\%), caused by validator defects rather than model behaviour; when those halts are counted as failed work, the governed arm scores below the ungoverned arm in four of six workflows. Protection therefore depends on correct contract and validator configuration, and halt attribution must itself be verified. The results support deterministic handoff validation as complementary to post-hoc failure attribution, learned guardrails, and agent-based verification; they do not establish universal correctness, hallucination detection, or model-independent effectiveness. We conclude with a model-dependence hypothesis and a cross-model validation protocol.
\end{abstract}

\begin{figure}[t]
\centering
\begin{tikzpicture}
\begin{axis}[
    ybar=4pt,
    bar width=10pt,
    width=\linewidth,
    height=6.5cm,
    ymin=0,
    ymax=7.8,
    ylabel={Mean rubric score (0--7)},
    symbolic x coords={B2B,Hospital,Dev,Enterprise,E-com,Insurance},
    xtick=data,
    x tick label style={rotate=25,anchor=east,font=\small},
    legend style={at={(0.5,1.02)},anchor=south,legend columns=-1,draw=none},
    enlarge x limits=0.09,
    clip=false
]

\addplot+[
    nodes near coords,
    every node near coord/.append style={font=\scriptsize,anchor=south,yshift=2pt,xshift=-2pt}
] coordinates {
    (B2B,5.06)
    (Hospital,6.00)
    (Dev,6.26)
    (Enterprise,6.50)
    (E-com,6.23)
    (Insurance,5.88)
};

\addplot+[
    nodes near coords,
    every node near coord/.append style={font=\scriptsize,anchor=south,yshift=2pt,xshift=2pt}
] coordinates {
    (B2B,6.53)
    (Hospital,6.67)
    (Dev,6.32)
    (Enterprise,7.00)
    (E-com,6.89)
    (Insurance,7.00)
};

\legend{Maat off,Maat intervene}
\end{axis}
\end{tikzpicture}
\caption{Mean deterministic rubric score on the paired subset (v2 scoring; see \cref{tab:mainresults} for $n$ per workflow). Descriptive means only.}
\label{fig:correctness}
\end{figure}
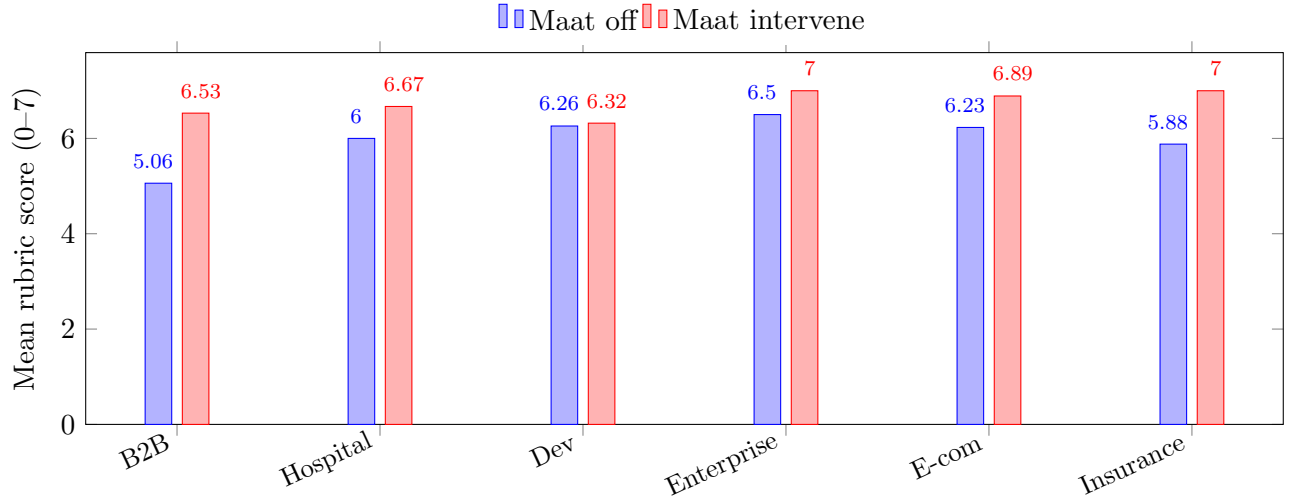

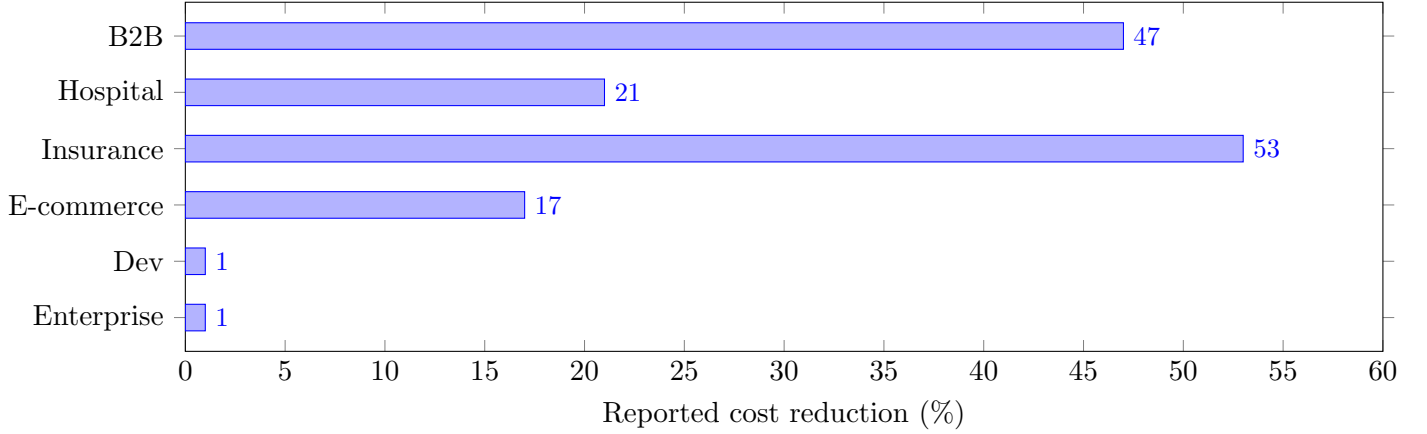
\begin{figure}[t]
\centering
\begin{tikzpicture}
\begin{axis}[
    xbar, width=\linewidth, height=6.2cm,
    xmin=0,xmax=60,
    xlabel={Reported cost reduction (\%)},
    symbolic y coords={Enterprise,Dev,E-commerce,Insurance,Hospital,B2B},
    ytick=data,
    nodes near coords, nodes near coords align={horizontal}, every node near coord/.append style={font=\small},
    enlarge y limits=0.12]
\addplot coordinates {(1,Enterprise) (1,Dev) (17,E-commerce) (53,Insurance) (21,Hospital) (47,B2B)};
\end{axis}
\end{tikzpicture}
\caption{Model-call/API cost reduction in the intervention arm on the paired subset (v2). Cost saving is produced by skipped downstream agents after a protective halt; it is not an intrinsic reduction in the cost of validation itself.}
\label{fig:cost}
\end{figure}

\noindent\fbox{\parbox{0.97\linewidth}{\small\textbf{Revision note (v2).} Version~1 stated that a protective halt was credited only when the associated finding corresponded to the injected defect. The deterministic scorers for the B2B, hospital-triage and software-development benchmarks did not implement this: any halt before the final artefact received full marks. The insurance and e-commerce scorers accepted any blocking finding as attribution. We re-scored every committed transcript without new model calls, hand-reviewed all 94 governed-arm halts, and replaced \cref{tab:mainresults} and \cref{fig:correctness,fig:cost} with corrected values. The qualitative conclusions on contract-expressible defects (tax, residency, identity, limits, price drift) are unchanged. The aggregate correctness and cost claims are substantially revised. The validator defects behind the false alarms, and the scorer defect, have since been fixed. The fixes were validated offline by replaying all saved trial outputs; a live re-run is pending and is not reported here.}}

\section{Introduction}
LLM-based multi-agent systems decompose complex tasks among specialized agents that exchange intermediate state, tool outputs, and decisions. This decomposition enables parallelism, specialization, context isolation, and structured collaboration, but it also creates additional interfaces at which errors can be introduced and propagated. MAST, a large empirical taxonomy of multi-agent failures, reports 14 failure modes across system-design, inter-agent-misalignment, and task-verification categories \citep{cemri2025mast}. Its current arXiv revision reports category shares of 44.2\%, 32.3\%, and 23.5\%, respectively. In parallel, controlled scaling experiments show architecture-dependent error amplification: systems without centralized verification can amplify trace-level errors substantially more than centrally coordinated systems \citep{kim2025scaling}. Recent work further studies error cascades \citep{xie2026spark,jamshidi2026cascade}, delayed verification \citep{itkin2026delay}, silent trajectory anomalies \citep{pathak2025silent}, and failure attribution \citep{kong2026aegis,liu2026masprism,qiao2026verifymas}.

These studies suggest that reliability is not solely a property of the base model. It is also a property of the coordination architecture and of the interfaces through which agents exchange state. Consider a sequential workflow with $n$ handoffs and independent per-stage success probability $p$. A simple lower-order model gives end-to-end success $R \approx p^n$, illustrating why small local error rates can become operationally material as interaction depth increases. Independence is not a realistic assumption for LLM-MAS errors may be correlated and can be amplified by downstream reuse, but the multiplicative model captures the basic interface-risk intuition.

Many proposed safeguards place a learned or LLM-based evaluator in the loop; agent-as-a-judge approaches, for example, use agentic systems to evaluate other agents \citep{zhuge2024agentjudge,you2026agentjudge}. Such evaluators are useful, but their verdicts are themselves probabilistic and can share failure modes with the systems they assess. Maat was designed around a narrow question: \emph{can a deterministic layer prevent contract-detectable handoff defects from propagating without asking another language model to judge them?} The system sits between agents and evaluates each handoff against a versioned contract describing what the workflow is authorized to do. The design intentionally separates generative reasoning from governance: agents may remain probabilistic, while the gate deciding whether a handoff violates an encoded rule is deterministic.

This paper makes four contributions:
\begin{enumerate}
    \item We formalize a deterministic contract-validation layer for multi-agent handoffs and describe its seven-gate architecture at the mechanism level.
    \item We report a controlled six-domain evaluation (522 trials) with deterministic, LLM-free scoring and distinguish end-to-end correctness from protective early termination.
    \item We identify \emph{configuration validity}, especially anchor-path resolution and entity normalization as a first-class reliability requirement, including a public correction to an earlier false-positive claim.
    \item We position deterministic handoff governance relative to attribution, anomaly detection, learned guardrails, and propagation-aware mitigation, and formulate a testable cross-model hypothesis rather than presenting it as an established result.
\end{enumerate}

\section{Problem Setting: Why Handoffs Matter}
A multi-agent system can fail even when every individual agent appears locally reasonable. Once an upstream agent emits a wrong value, omitted field, or inconsistent identity, downstream agents may treat that state as authoritative and construct increasingly coherent output around it. The benchmark suite was designed around this propagation pattern: discount fabrication becomes an incorrect invoice, patient-identity drift becomes a downstream safety and privacy risk, and a wrong insurance amount can become a formally explained but contractually invalid payout.

The elementary reliability calculation $p^n$ is useful only as intuition: if each of $n$ sequential stages succeeds with probability $p$ and failures were independent, end-to-end reliability would decrease multiplicatively with workflow depth. Real LLM systems violate the independence assumption because downstream agents reuse upstream state; an early defect can therefore be amplified rather than merely accumulated. This motivates placing verification at the seam between agents instead of evaluating only the final answer.

Three distinctions are important throughout the paper. First, \emph{structural validity} is not the same as \emph{contract validity}: a payload may be well formed and still violate a price, payout, tax, or authorization rule. Second, \emph{determinism} is not the same as \emph{correctness}: a wrongly configured deterministic rule can reproducibly produce the wrong verdict. Third, \emph{detectability} is not the same as \emph{universality}: Maat can reject only defects that become observable as encoded handoff or contract violations.

\section{Related Work}
\paragraph{Failure taxonomies and attribution.}
MAST provides an empirical taxonomy and annotated dataset for understanding why multi-agent systems fail \citep{cemri2025mast}. Aegis scales failure-attribution research through automated, context-aware error generation and a 9,533-trajectory dataset \citep{kong2026aegis}. MASPrism uses prefill-stage signals for lightweight failure attribution \citep{liu2026masprism}, while VerifyMAS frames attribution as hypothesis verification \citep{qiao2026verifymas}. These methods primarily answer \emph{where did the failure originate?} Maat instead acts before the receiving agent consumes a handoff and asks \emph{does this transition satisfy an explicit contract?}

\paragraph{Trajectory anomalies and guardrails.}
Pathak et al. study silent failures such as drift, cycles, and missing details as anomaly detection over multi-agent trajectories \citep{pathak2025silent}. Bi-level graph anomaly detection has also been proposed for fine-grained safeguarding \citep{pan2025xguard}. AgentDoG develops an agentic safety taxonomy and a learned diagnostic guardrail \citep{agentdog2026}. These approaches broaden detectable behavior beyond explicit contracts, but their learned or statistical components are probabilistic. Maat makes a narrower guarantee: for a fixed engine version, configuration, anchor, and handoff, the verdict is reproducible.

\paragraph{Propagation-aware safety.}
From Spark to Fire models error propagation as a dependency graph and proposes a genealogy-based governance layer \citep{xie2026spark}; PropGuard similarly targets propagation-aware exploration and remediation \citep{yan2026propguard}. Hallucination Cascade studies claim-level evolution across sequential agents \citep{jamshidi2026cascade}. Itkin analyzes delayed verifier dynamics and shows that verification timing itself can affect stability in belief-update systems \citep{itkin2026delay}. Maat is complementary: it does not model semantic belief dynamics; it blocks explicit structural and contract violations at the interface where they become observable.

\paragraph{Architecture and shared state.}
Kim et al. isolate coordination architecture as a causal experimental factor and report that centralized verification bottlenecks reduce trace-level error amplification \citep{kim2025scaling}. Agora, in a different setting, shows the value of durable shared state, lineage, and reproducibility for coordinating autonomous research agents \citep{zhang2026agora}. These results support a broader systems view in which reliability depends on external coordination state, not only on individual model capability.

\section{Maat}
\subsection{Public system model and disclosure boundary}
At the public-interface level, a multi-agent workflow can be represented as a directed graph $G=(V,E)$ whose nodes are agents and whose edges are handoffs. Each handoff carries a structured payload together with execution metadata and relevant workflow state. Maat is inserted on selected edges and compares the observed handoff with a versioned \emph{anchor}: the explicit state that the workflow is authorized to preserve or obey, such as contractual limits, approved entities, required fields, jurisdictional constraints, or policy terms.

For a fixed engine version, fixed anchor, fixed public configuration, and identical handoff state, the validator returns the same verdict. This deterministic input--output property is the reproducibility claim evaluated in this paper. We intentionally do not publish the conductor's internal evaluation order, rule-compilation procedure, configuration-construction methodology, or implementation-specific state model. Those elements are not required to evaluate the behavioral claim and constitute part of the proprietary implementation. The formalism in this article therefore specifies the \emph{observable contract of the validator}, not its internal algorithm.

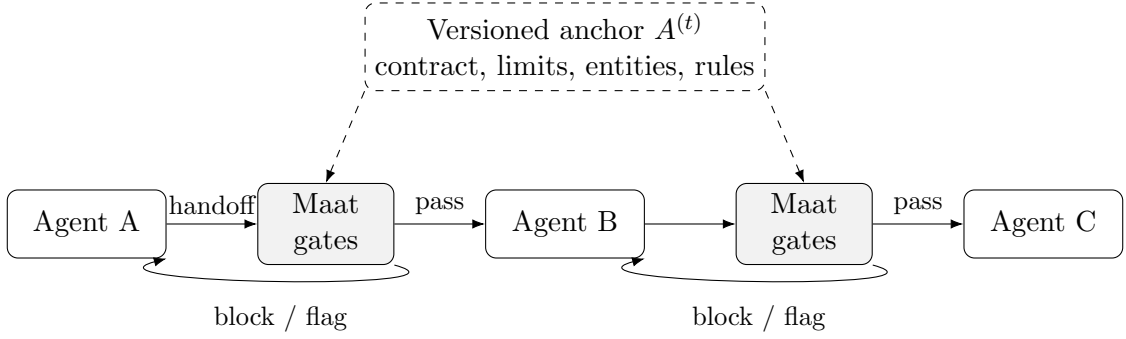
\begin{figure}[t]
\centering
\begin{tikzpicture}[
    node distance=9mm and 12mm,
    >=Latex,
    agent/.style={draw,rounded corners,minimum width=21mm,minimum height=9mm,align=center},
    gate/.style={draw,rounded corners,minimum width=18mm,minimum height=9mm,align=center,fill=gray!10},
    anchorbox/.style={draw,dashed,rounded corners,minimum width=36mm,minimum height=10mm,align=center}
]

\node[agent] (a) {Agent A};
\node[gate,right=of a] (g1) {Maat\\gates};
\node[agent,right=of g1] (b) {Agent B};
\node[gate,right=of b] (g2) {Maat\\gates};
\node[agent,right=of g2] (c) {Agent C};
\node[anchorbox,above=13mm of b] (an)
    {Versioned anchor $A^{(t)}$\\contract, limits, entities, rules};

\draw[->] (a.east) -- node[above,font=\small]{handoff} (g1.west);
\draw[->] (g1.east) -- node[above,font=\small]{pass} (b.west);
\draw[->] (b.east) -- (g2.west);
\draw[->] (g2.east) -- node[above,font=\small]{pass} (c.west);

\draw[->,dashed] (an.south west) -- (g1.north);
\draw[->,dashed] (an.south east) -- (g2.north);

\draw[->]
    (g1.south east)
    to[out=-15,in=-165,looseness=1.12]
    node[pos=.50,below=5pt,font=\small]{block / flag}
    (a.south east);

\draw[->]
    (g2.south east)
    to[out=-15,in=-165,looseness=1.12]
    node[pos=.50,below=5pt,font=\small]{block / flag}
    (b.south east);

\end{tikzpicture}
\caption{Maat is inserted at agent boundaries. Each handoff is checked against the same versioned anchor before the downstream agent consumes it. The validation path contains no language model.}
\label{fig:architecture}
\end{figure}

\subsection{Seven gates and requirement primitives}
The public architecture comprises seven gates: (G1) plan validation; (G2) handoff validation; (G3) validation-depth control; (G4) role adherence and integrity; (G5) survival/health monitoring; (G6) a circuit breaker for persistently failing agents; and (G7) cross-handoff entity consistency. Requirement primitives sit above the structural gates and encode domain rules such as payout limits, permitted discounts, tax treatment, provider eligibility, or data-residency constraints.

The distinction is important. Structural gates determine whether the workflow and handoff are well formed; requirement primitives determine whether a well-formed handoff is allowed by the governing contract. A syntactically valid value can still be contractually wrong. Conversely, Maat does not claim to detect every hallucination: a semantically false statement can pass if it does not violate an encoded structural or contract rule.

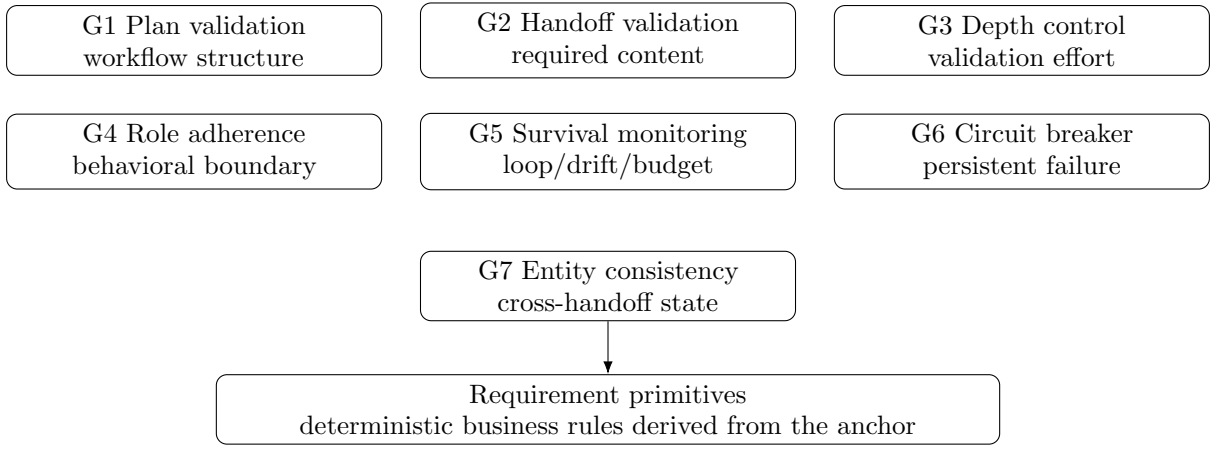
\begin{figure}[t]
\centering
\begin{tikzpicture}[
    node distance=5mm and 5mm,
    >=Latex,
    every node/.style={font=\small},
    box/.style={draw,rounded corners,minimum height=8mm,align=center,text width=0.27\linewidth}
]

\node[box] (g1) {G1 Plan validation\\workflow structure};
\node[box,right=of g1] (g2) {G2 Handoff validation\\required content};
\node[box,right=of g2] (g3) {G3 Depth control\\validation effort};

\node[box,below=of g1] (g4) {G4 Role adherence\\behavioral boundary};
\node[box,right=of g4] (g5) {G5 Survival monitoring\\loop/drift/budget};
\node[box,right=of g5] (g6) {G6 Circuit breaker\\persistent failure};

\node[box,below=8mm of g5] (g7) {G7 Entity consistency\\cross-handoff state};

\node[box,below=7mm of g7,text width=0.58\linewidth] (req)
    {Requirement primitives\\deterministic business rules derived from the anchor};

\draw[->] (g7.south) -- (req.north);

\end{tikzpicture}
\caption{Mechanism-level decomposition. G1--G7 constitute the public gate layer. Requirement primitives complement these gates with deterministic business rules derived from the anchor. The domain benchmarks principally exercise G2, G7, and requirement primitives; separate public tests stress G3, G5, and G6.}
\label{fig:gates}
\end{figure}

\subsection{Framework integration artifact: CrewAI}
To test whether the boundary mechanism can be inserted into an existing agent framework without rewriting the workflow, we implemented a CrewAI integration artifact with three layers. First, a local structural chain analyzer introspects a CrewAI \texttt{Flow} and deterministically reports wiring defects such as dangling edges, unreachable steps, and orphaned starts; optional \texttt{produces}/\texttt{needs} annotations enable field-level unmet-requirement checks. Second, a lightweight runtime connector inserts \texttt{validate()} at a handoff and performs local structural checks when no contract is configured, or calls the Maat conductor when a registered plan and anchor are supplied. Third, a minimal intake--pricing--invoice demo exercises a business-rule violation in which a structurally valid pricing payload applies a 40\% discount although the registered contract authorizes only 10\%. The ungoverned path accepts the arithmetic and issues the discounted invoice; the governed path rejects the \texttt{pricing\_to\_invoice} handoff with \texttt{REQ\_DISCOUNT\_EXCEEDED}.

This artifact is an integration demonstration rather than an additional benchmark arm. It establishes that the mechanism can operate at a concrete framework boundary and illustrates the distinction between \emph{structural validation} and \emph{contract validation}. Without a plan and anchor, Maat can check shape, presence, and provenance-related properties but cannot infer workflow-specific commercial limits. With the anchor configured, the same handoff can be evaluated against authorized business state. This distinction is consistent with the configuration-dependence observed in the benchmark suite and makes explicit where domain knowledge enters the system.

\subsection{Configuration as part of the system}
A deterministic rule is useful only if it evaluates the intended state. Two benchmark episodes exposed the same configuration failure class: enabled rules did not correspond to the scenario state that the benchmark intended them to judge. In that state, the engine remained deterministic yet produced ineffective or false findings. This leads to a stronger system definition: the evaluated object is not the validator binary alone but the deployed combination of engine version, anchor version, and configuration state. The exact construction procedure for that configuration is intentionally outside the public paper; what the experiments show is that it must itself be tested and versioned.

Two public engineering lessons follow. First, a declared rule should be exercised against representative live-shaped payloads before validator silence is interpreted as evidence of correctness. Second, comparisons over collections should not treat semantically irrelevant reordering as a change in business state. These are observable correctness requirements; they do not disclose the proprietary rule-construction methodology.

\section{Experimental Method}
\subsection{Research questions}
We evaluate four questions:
\begin{description}
\item[RQ1:] Does deterministic handoff validation improve end-to-end contract correctness on controlled multi-agent workflows?
\item[RQ2:] When intervention stops a defective chain, how does model-call cost change relative to the ungoverned workflow?
\item[RQ3:] Which concrete failure modes are intercepted, and which remain outside the validated coverage?
\item[RQ4:] How sensitive are results to anchor and rule configuration?
\end{description}

\subsection{Three-arm design}
Each domain benchmark is executed in three modes: \emph{off} (unguarded baseline), \emph{warn} (findings logged but execution continues), and \emph{intervene} (a blocking finding terminates or redirects the chain). The reported cross-benchmark correctness and cost comparison uses the off and intervene arms. Failure profiles corrupt data or context that an agent treats as authoritative rather than merely instructing an agent to ``be careless.'' This choice matters because capable models often resist an explicit instruction to make an error; the benchmark is intended to study error propagation after bad state enters the workflow, not willingness to follow an obviously adversarial prompt.

The public methodology uses a deterministic seven-check rubric, normalized to a 0--7 score, with no LLM judge in the scoring path. In the corrected scoring (v2), a halt receives full marks only when (i) the benchmark's defect-present check confirms that the injected defect had entered the chain at or before the halting handoff, and (ii) the blocking finding is a specific finding rather than the generic ``no provenance'' signal. In the B2B workflow, a halt on a verified arithmetic inconsistency in the deal record (total $\neq$ sum of line items) is also credited, even though that error was not injected. Any other halt is unattributed. Version~1 did not apply this rule in three of six scorers (see the revision note). The benchmark draft identifies Claude Haiku 4.5 as the single base model used across the domain runs. Consequently, model variation is not part of the present experiment.

\begin{figure}[t]
\centering
\begin{tikzpicture}[node distance=6mm and 8mm,>=Latex,
small/.style={draw,rounded corners,align=center,minimum height=8mm,text width=0.21\linewidth},
box/.style={draw,rounded corners,align=center,minimum height=8mm,text width=0.25\linewidth}]
\node[small] (input) {Same seed, anchor,\\failure profile};
\node[box,right=of input] (off) {OFF\\agents run unchecked};
\node[box,below=of off] (warn) {WARN\\validate and log; continue};
\node[box,below=of warn] (int) {INTERVENE\\block on configured finding};
\draw[->] (input)--(off);
\draw[->] (input.east) -- ++(4mm,0) |- (warn.west);
\draw[->] (input.east) -- ++(4mm,0) |- (int.west);
\end{tikzpicture}
\caption{Controlled three-arm design. Workflow inputs and injected defects are held fixed while the governance posture changes.}
\label{fig:arms}
\end{figure}
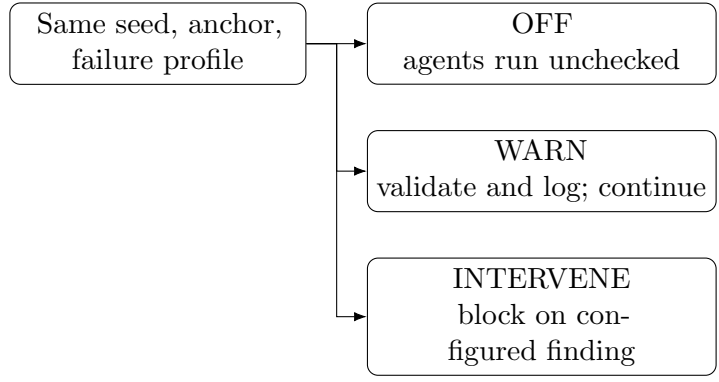

\subsection{Domain workflows}
The six workflows are summarized in \cref{tab:benchmarks}. Trial counts sum to 522. Agent count follows the task decomposition rather than a fixed synthetic team size.

\begin{table}[t]
\centering
\caption{Domain benchmarks.}
\label{tab:benchmarks}
\small
\begin{tabularx}{\linewidth}{lrrX}
\toprule
Benchmark & Agents & Trials & Workflow \\
\midrule
B2B financial & 6 & 75 & lead-to-invoice \\
Hospital triage & 10 & 75 & intake-to-disposition \\
Software development & 12 & 75 & specification-to-release \\
Enterprise discovery & 12 & 90 & process discovery with amendments \\
E-commerce agency & 15 & 135 & multi-marketplace, two concurrent clients \\
Health insurance & 10 & 72 & EU claims adjudication \\
\midrule
Total & & 522 & \\
\bottomrule
\end{tabularx}
\end{table}

\subsection{What each benchmark is intended to stress}
The six domains exercise different seam-level failure patterns rather than serving as six claims of production readiness.
\begin{itemize}
    \item \textbf{B2B financial, lead-to-invoice:} deal terms, usage quantities, service tiers, and discounts must survive a largely linear chain without unauthorized drift.
    \item \textbf{Hospital triage, intake-to-disposition:} patient identity, allergy state, medication facts, and risk flags must remain coherent across ten roles. This is a controlled safety-relevant simulation, not a clinical evaluation.
    \item \textbf{Software development, specification-to-release:} parallel branches and repeated access to the original specification create natural redundancy, making this a useful lower-bound case for the incremental value of an external gate.
    \item \textbf{Enterprise process discovery:} parallel source scanners and authorized mid-flight amendments stress scope changes, legal-approval state, and continued propagation of the amended authority.
    \item \textbf{E-commerce agency:} a 15-agent, two-client, multi-marketplace workflow stresses cross-client contamination, marketplace price drift, VAT treatment, revenue reconciliation, and contract scope under concurrency.
    \item \textbf{Health insurance adjudication:} the most contract-dense workflow stresses claimant identity, coverage, payout logic, provider eligibility, jurisdiction, tax treatment, data residency, and scenario-conditioned expected values.
\end{itemize}

\subsection{Result versioning}
The benchmark evidence evolved between the initial July 2026 test runs and the public repository snapshot used for this September 2026 article. The later repository snapshot revises the health-insurance row and documents 11 false-positive findings that were not present in the earlier internal benchmark snapshot. Because this article is intended to represent the current public evidence, \cref{tab:mainresults} uses the later repository values. This revision is consistent with the paper's central methodological claim: configuration state is part of the measured system and must be versioned with the result. Version~2 adds a second, larger revision: a scorer audit and a manual review of every governed-arm halt (\cref{sec:fpaudit}).

\section{Results}
\subsection{End-to-end correctness and cost}
\begin{table}[t]
\centering
\caption{Corrected cross-benchmark results (v2). \emph{v1} reproduces the originally published means. \emph{Paired} keeps only (profile, seed) pairs in which the governed run completed or halted on an attributable finding; $n$ is the number of pairs kept out of the governed trials. \emph{Lower bound} scores every unattributed halt as delivering no correct output. Insurance uses a scenario-conditioned payout key (correct denials expect EUR~0). Percentages are descriptive changes, not inferential estimates.}
\label{tab:mainresults}
\small
\setlength{\tabcolsep}{4pt}
\begin{tabular}{lrrrrrrr}
\toprule
 & \multicolumn{2}{c}{$\Delta$ correctness} & & & & \multicolumn{2}{c}{Cost $\Delta$} \\
\cmidrule(lr){2-3}\cmidrule(lr){7-8}
Benchmark & v1 & Paired & $n$ & Lower bound & Halts (unattr.) & v1 & Paired \\
\midrule
B2B financial & +26.5\% & +29.1\% & 17/25 & -15.9\% & 20 (8) & -50\% & -47\% \\
Hospital triage & +12.4\% & +11.1\% & 9/25 & -60.8\% & 22 (16) & -49\% & -21\% \\
Software development & +3.8\% & +0.8\% & 19/25 & -23.1\% & 7 (6) & -8\% & -1\% \\
Enterprise discovery & +7.7\% & +7.7\% & 30/30 & +7.7\% & 5 (0) & $\sim$0\% & $\sim$0\% \\
E-commerce agency & +10.7\% & +10.6\% & 44/45 & +8.3\% & 18 (1) & -17\% & -17\% \\
Health insurance$^{*}$ & +2.9\% & +19.1\% & 16/24 & -21.1\% & 22 (8) & -55\% & -53\% \\
\bottomrule
\end{tabular}
\vspace{1mm}
\begin{minipage}{0.96\linewidth}\footnotesize
$^{*}$With the scenario-conditioned payout key, the full insurance grid is 5.92 (off) vs.\ 5.83 (intervene), i.e.\ $-1.4\%$. The v1 value of +2.9\% used an unconditioned key that penalised correct denials in the ungoverned arm. The paired insurance subset excludes both deny profiles, so its +19.1\% should not be read as a whole-benchmark effect.
\end{minipage}
\end{table}

On the paired subset, five of six workflows improve and software development is flat. The B2B gain rests on 17 pairs, of which 8 halts stopped a genuine but un-injected arithmetic error in the deal record. The hospital gain rests on only 9 pairs, because 16 of its 22 halts were unattributed. The lower-bound column shows why attribution matters: in four workflows the governed arm falls below the ungoverned arm once false-alarm halts are counted as failed work. The version~1 claim that intervention improves correctness on every workflow is therefore withdrawn. The supported claim is narrower: where halts are attributable, they prevent the defect from propagating.

The cost effect is conditional rather than universal. Intervention reduces cost when a blocking defect occurs early enough to avoid downstream model calls. Enterprise discovery is approximately cost-neutral because its measured defects and workflow structure do not consistently remove enough downstream work. Thus, Maat should not be described as a generic cost-reduction mechanism; it is an early-termination mechanism whose cost effect depends on defect prevalence and location. In version~1, much of the hospital and B2B saving came from false-alarm halts, which save cost only by stopping correct work.

\subsection{Representative contract catches}
The public benchmark record includes several failures for which the intended rule or entity-consistency mechanism fired:
\begin{itemize}
    \item \textbf{Cross-marketplace price drift:} the same SKU appeared at EUR 27.99, EUR 21.99, and EUR 19.99 across three marketplaces. A keyed entity-consistency rule detected the divergence before further propagation.
    \item \textbf{VAT/tax misclassification:} jurisdiction-dependent tax treatment was caught in e-commerce and insurance scenarios when the defect manifested.
    \item \textbf{Payout inflation:} an insurance scenario propagated EUR 1,310 where the contract-correct payout was EUR 810; a value-mismatch rule halted the governed chain before payment propagation.
    \item \textbf{Claimant identity drift:} a changed policyholder identifier was detected through cross-handoff entity tracking.
    \item \textbf{Data residency:} a scenario marking special-category medical data for non-EU processing triggered a configured residency rule.
    \item \textbf{Un-injected revenue mis-report:} an e-commerce analytics agent repeatedly double-counted marketplace revenue without the benchmark explicitly injecting that particular error; in version~1 the configured validator recorded it only as a warning and did not stop the chain (it is a blocking rule since v2).
\end{itemize}
These examples demonstrate contract and state consistency, not open-domain factuality. A fluent but wrong statement with no corresponding anchor rule remains outside the mechanism's guaranteed coverage.

\begin{table}[t]
\centering
\caption{Illustrative defects and the observable property that made them detectable.}
\small
\begin{tabularx}{\linewidth}{lXX}
\toprule
Case & Observable violation & Why the boundary is checkable \\
\midrule
Marketplace price drift & Same keyed SKU carries conflicting prices & Stable entity, divergent tracked value \\
Payout inflation & Published payout differs from contract-derived expected value & Numeric decision conflicts with encoded authority \\
VAT/tax mismatch & Transaction classification conflicts with jurisdiction rule & Jurisdiction and transaction class are explicit state \\
Claimant identity drift & Policyholder identifier changes across handoffs & Identity should remain invariant through adjudication \\
Data residency & Processing destination violates the encoded rule & Processing location is explicit policy-bounded state \\
Revenue mis-report & Aggregate does not reconcile to source transactions & Deterministic reconciliation exposes the mismatch \\
\bottomrule
\end{tabularx}
\end{table}

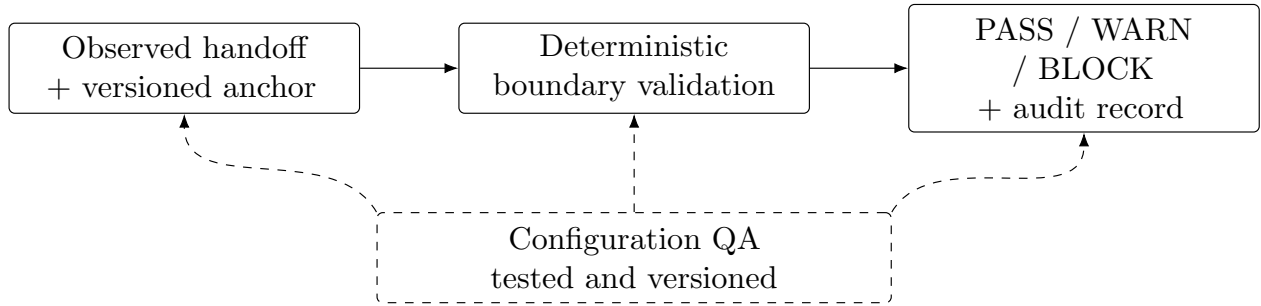
\begin{figure}[t]
\centering
\vspace{2mm}
\resizebox{0.95\linewidth}{!}{%
\begin{tikzpicture}[
    >=Latex,
    node distance=10mm and 12mm,
    box/.style={
        draw,
        rounded corners=2pt,
        align=center,
        minimum height=11mm,
        text width=40mm,
        inner sep=4pt
    },
    qa/.style={
        draw,
        dashed,
        rounded corners=2pt,
        align=center,
        minimum height=11mm,
        text width=60mm,
        inner sep=4pt
    }
]
\node[box] (state) {Observed handoff\\+ versioned anchor};
\node[box, right=of state] (val) {Deterministic\\boundary validation};
\node[box, right=of val] (out) {PASS / WARN / BLOCK\\+ audit record};

\node[qa, below=12mm of val] (config)
{Configuration QA\\tested and versioned};

\draw[->] (state.east) -- (val.west);
\draw[->] (val.east) -- (out.west);

\draw[->, dashed] (config.north west)
    to[out=120,in=-90] (state.south);
\draw[->, dashed] (config.north)
    -- (val.south);
\draw[->, dashed] (config.north east)
    to[out=60,in=-90] (out.south);
\end{tikzpicture}%
}
\vspace{2mm}
\caption{Public configuration principle. Deterministic execution is reproducible only with respect to the deployed anchor and configuration; those artifacts must therefore be tested and versioned with the engine. Internal rule construction is outside the disclosure boundary of this paper.}
\label{fig:config}
\end{figure}

\subsection{Configuration correction and false positives}
\label{sec:fpaudit}
\paragraph{Halt audit (v2).} We read every one of the 94 halts in the intervention arms of the six workflows. In 45 the halt stopped the injected defect. In 13 it stopped a genuine defect that was not injected: deal totals inconsistent with line items, a missing required field, and an out-of-network payout. One was a legitimate escalation, where the agent could not determine a payout. The remaining 35 (37\%) were false alarms on correct output: 17 of 22 halts in hospital triage, 8 of 20 in B2B, 6 of 22 in insurance, 3 of 7 in software development, 1 of 18 in e-commerce and 0 of 5 in enterprise discovery. The causes were validator defects, not model behaviour:
\begin{itemize}
\item (22 halts) Circuit-breaker state carried over from earlier trials escalated validation to its strictest depth. At that depth provenance became mandatory, and the provenance check recognised only a top-level key, so it ignored field-level and per-item provenance that the agents had supplied.
\item (8) A legitimately empty answer, such as no declared allergies or no documented history, was treated as a hollow handoff.
\item (4) Correct EUR~0 denials were compared with a payable-claim expected amount.
\item (1) Per-country transaction counts were parsed as tax rates.
\end{itemize}
Because the benchmark scorers credited these halts, the version~1 table counted them as successes.

The strongest negative result is also the most operationally useful. An earlier internal benchmark snapshot reported zero false positives on the deterministic rubric. The later public repository documents 11 false-positive findings in the health-insurance benchmark, concentrated in \texttt{REQ\_VALUE\_MISMATCH} and \texttt{INFO\_EMPTY} on two profiles where the model correctly denied a claim while the anchor's expected payout still represented the clean-claim case. This is not model nondeterminism: it is a mismatch between the validator's expected value and the scenario state.

Accordingly, determinism should not be confused with correctness. Determinism means that the same misconfigured rule will fail in the same way every time. The practical benefit is diagnosability and reproducibility; the practical burden is that anchor construction and scenario-conditioned expected values must themselves be tested.

\subsection{Behavioral gates}
Separate public tests stress mechanisms not naturally exercised by data-corruption domain scenarios. The reported circuit-breaker benchmark shows 0\% false trips on a healthy pipeline, 100\% trips under an engineered chronic-failure profile, and approximately 37\% lower cost than running the chronically failing agent without the breaker. A validation-depth sweep reports four additional findings per trial under thorough versus light validation and a lower aggregate cost on defective traffic because earlier halts skip downstream calls. Survival monitoring is characterized as calibration-dependent rather than governed by a universal threshold. These results should be read as targeted mechanism tests; they do not establish that every gate is equally validated across the six domain workflows.

\section{Operational Scope: Where Maat Is and Is Not a Good Fit}
The experiments suggest that deterministic boundary governance is most useful when a workflow carries a stable, inspectable source of authority and when expensive or regulated consequences occur downstream of the violating handoff. Examples include insurance policies, contractual price limits, financial authorizations, tax rules, data-residency requirements, approved identities, clinical protocol constraints, and service-level obligations.

It is less compelling when downstream agents already possess strong independent redundancy, when there is no stable contract to encode, or when the target failure is purely semantic and cannot be reduced to a checkable invariant. The software-development benchmark illustrates the first case: agents frequently re-derive upstream requirements, so external governance adds less. A hallucinated explanation that violates no encoded state illustrates the second: the statement may be false yet still pass a contract layer.

\begin{table}[t]
\centering
\caption{Practical fit of deterministic handoff governance.}
\small
\begin{tabularx}{\linewidth}{XX}
\toprule
Stronger fit & Weaker fit \\
\midrule
Stable policy or contract exists & No authoritative state can be encoded \\
Identity/value must survive many handoffs & Agents already independently reconstruct the same facts \\
Violation has financial, legal, safety, or audit impact & Errors are harmless or cheaply reversible \\
Downstream model calls are expensive & Workflow is short and inexpensive \\
Auditability and reproducibility matter & Open-ended semantic quality is the main objective \\
\bottomrule
\end{tabularx}
\end{table}

\subsection{Deployment posture}
The public design supports two operational postures. In \emph{warn mode}, selected handoffs are checked and findings are recorded while the workflow continues. This is the appropriate starting posture for a new anchor because the organization can observe what would have been blocked while configuration is calibrated. In \emph{intervene mode}, findings designated as blocking halt the chain or route execution to recovery. Only intervention can create the measured downstream cost saving because only intervention prevents later model calls after a detected defect. Routing a blocked handoff to a human reviewer, rather than letting the chain continue or silently repairing the payload, is one way to support the human-oversight requirements that the EU AI Act sets for high-risk systems \citep{eu2024aiact}; the validator on its own does not constitute compliance.

\section{Discussion}
\subsection{What the results support}
The benchmark suite supports a limited but meaningful conclusion: when a defect is expressible as a deterministic predicate over the current handoff, prior tracked state, and a correctly configured contract, inline validation can prevent that defect from reaching later agents. This is especially useful for quantities and identities expected to remain stable across handoffs, explicit eligibility rules, limits, and jurisdictional constraints.

This observation aligns with architecture-level evidence that validation bottlenecks can reduce error propagation \citep{kim2025scaling}. It also complements attribution work \citep{kong2026aegis,liu2026masprism,qiao2026verifymas}: attribution explains a failed trajectory after the fact, whereas contract gating can stop some trajectories before the downstream effect occurs. The two functions are not substitutes.

\subsection{What the results do not support}
First, the experiments do not establish general hallucination detection. Maat cannot reject an incorrect assertion simply because it is false unless falsity manifests as a configured contract or consistency violation. Second, the evaluation uses a single base model and therefore cannot establish model-independent utility. Third, observed benchmark improvements are descriptive; the present public result tables do not provide the per-trial distributions needed for inferential claims. Fourth, the workflows are controlled simulations of operational domains, not prospective deployments inside hospitals, insurers, or production marketplaces.

Finally, the false-positive revision shows that an audit-friendly deterministic layer can still be wrong because its specification is wrong. This shifts part of the engineering problem from ``how do we judge agent output?'' to ``how do we validate the contract representation that performs the judging?'' We view that shift as desirable because specification errors are inspectable, testable, and versionable, but it is not free.

\subsection{Model-dependence hypothesis}
Several injected defects in the source experiments failed to manifest because the base model self-corrected. We therefore propose, but do not claim to have demonstrated, the following hypothesis:

\begin{quote}
For defects that are representable by a fixed deterministic contract, the validator's verdict is invariant to the model used by the producing agent, while the \emph{frequency with which such defects manifest} is model-dependent.
\end{quote}

A direct test is a paired cross-model experiment: hold workflow, seeds, anchor, injections, and validator version fixed; replace only the base model with a stronger and a cost-optimized model; measure (i) defect manifestation rate, (ii) validator recall conditional on manifestation, (iii) false-positive findings, and (iv) total cost. The prediction is not that weaker models make every defect more often, but that the deterministic conditional detection behavior remains stable once the same violating handoff occurs.

\section{Limitations and Threats to Validity}
\paragraph{Single-model evaluation.} The current domain results use one base model. Cross-model generalization remains open.

\paragraph{Synthetic but structured workflows.} The benchmarks are designed to resemble real operational chains but do not reproduce the full organizational, legal, data-quality, and latency conditions of production deployments.

\paragraph{Scorer attribution.} Version~1 over-credited halts (see \cref{sec:fpaudit}). The corrected paired estimates rest on small $n$ for B2B (17) and hospital (9). The B2B anchor encodes a single customer's contract for all five customer seeds, so anchor-based B2B checks are only meaningful for one seed.

\paragraph{Configuration dependence.} The validator is only as accurate as the anchor and rule bindings. The documented insurance false positives make this a measured limitation rather than a theoretical concern.

\paragraph{Coverage.} Public evidence is strongest for handoff completeness, requirement primitives, entity consistency, validation depth, survival calibration, and circuit breaking. The public results do not justify claiming equivalent empirical validation of every declared gate.

\paragraph{No inferential statistics.} The currently published summaries provide aggregate means and profile outcomes. Without complete per-trial distributions in the article package, we intentionally avoid $p$-values, confidence intervals, and significance language.

\paragraph{Cost portability.} API prices and token usage patterns vary by provider, model, prompt length, caching, and deployment. The reported dollar values are benchmark-specific; the portable mechanism is skipped downstream inference after early termination.

\section{Reproducibility and Artifact Availability}
Benchmark methodology and public result files are available at \url{https://github.com/Lorelys/maat-benchmarks}. The project page is \url{https://maat.synwe.ai/}. The public benchmark repository reports benchmark designs, per-benchmark summaries, gate analyses, and consolidated tables. In addition, a CrewAI integration artifact provides a local deterministic structural analyzer, a handoff connector, a recorded verdict replay, and a minimal business-rule demo. The artifact does not expose the proprietary conductor implementation or the complete configuration methodology, so third parties can inspect the integration behavior and reproduce the local structural analysis, but cannot independently reproduce every end-to-end conductor decision from source alone. We therefore distinguish \emph{artifact-level reproducibility} of the open analyzer/demo from \emph{full implementation reproducibility} of the proprietary runtime. This disclosure boundary is deliberate: the paper publishes enough information to evaluate the scientific claim without publishing the internal configuration methodology identified as licensable intellectual property.

\section{Conclusion}
We presented Maat, a deterministic contract-based governance layer for multi-agent LLM workflows. Across six controlled domain benchmarks, corrected results show improvements where halts are attributable to real defects (five of six workflows on the paired subset). A hand review, however, found that 37\% of version~1 halts were false alarms caused by validator defects, and the originally reported gains on every workflow do not survive attribution-aware scoring. More importantly, the benchmark history exposes the central engineering constraint: determinism does not remove specification risk. Anchor construction, path resolution, expected-value semantics, and entity normalization are part of the system and must be validated and versioned with the result. Maat is therefore best understood not as a universal AI judge, but as an external deterministic boundary for the subset of agent behavior that can be made explicit as contracts and state invariants. The next decisive experiment is cross-model replication with fixed contracts and identical defect profiles.

\bibliographystyle{plainnat}
\bibliography{references}

\appendix
\section{Mapping to Selected MAST Failure Modes}
The mapping in \cref{tab:mastmap} is conceptual coverage, not a claim that all MAST modes are fully solved. Frequencies are omitted because the latest MAST revision differs from percentages used in earlier internal drafts; the paper uses the current category-level figures in the Introduction and avoids mixing revision-specific mode frequencies.

\begin{table}[h]
\centering
\caption{Conceptual mapping between selected MAST failure modes and Maat mechanisms.}
\label{tab:mastmap}
\small
\begin{tabularx}{\linewidth}{lX}
\toprule
MAST failure mode & Maat mechanism \\
\midrule
Disobey task specification & G2 plus requirement primitives compare output to the anchor \\
Disobey role specification & G4 checks declared role boundaries (public empirical coverage remains limited) \\
Information withholding & G2 required-field completeness \\
Ignored other agent's input & G7 cross-handoff entity consistency \\
Premature termination / unhealthy execution & G5--G6 survival monitoring and circuit breaking \\
No or incomplete verification & deterministic validation at handoff boundaries \\
\bottomrule
\end{tabularx}
\end{table}

\section{Benchmark Interpretation Checklist}
For each reported gate or primitive, the following states should be distinguished before drawing a conclusion from a non-firing validator:
\begin{enumerate}
    \item \textbf{Not exercised:} the benchmark never creates the corresponding defect.
    \item \textbf{Pre-empted:} an earlier blocking finding prevents the workflow from reaching the intended check.
    \item \textbf{Self-corrected:} the agent repairs or rejects the injected corruption, so no violating handoff is emitted.
    \item \textbf{Misconfigured:} the rule is enabled but reads the wrong field, wrong scenario expectation, or wrong entity representation.
    \item \textbf{True negative:} the relevant handoff is valid and the rule correctly remains silent.
\end{enumerate}
This taxonomy prevents ``no alert'' from being conflated with ``validated correct.''

\section{Caught Scenarios and Handoff Locations}
\label{app:caught}
The benchmark trace record identifies not only whether a defect was detected, but also the producer whose output was validated, the downstream consumer that would otherwise have received that output, the handoff at which validation occurred, and the gate or requirement primitive responsible for the finding. This appendix reports representative caught scenarios to make the interception points explicit. It is trace-level provenance rather than a separate benchmark: aggregate results remain those reported in \cref{tab:mainresults}.

\begin{table}[h]
\centering
\caption{Representative caught scenarios and their interception boundaries.}
\label{tab:caughtscenarios}
\scriptsize
\begin{tabularx}{\linewidth}{>{\raggedright\arraybackslash}p{0.11\linewidth} >{\raggedright\arraybackslash}p{0.18\linewidth} >{\raggedright\arraybackslash}p{0.22\linewidth} >{\raggedright\arraybackslash}p{0.24\linewidth} X}
\toprule
Benchmark & Scenario & Mechanism & Producer $\rightarrow$ consumer & Handoff \\
\midrule
Insurance & Payout inflation & \texttt{REQ\_VALUE\_MISMATCH} & decision $\rightarrow$ payment & decision/payment \\
Insurance & Claimant identity drift & G7 entity drift & coverage $\rightarrow$ medical & coverage/medical \\
Insurance & Out-of-network provider & \texttt{REQ\_PROVIDER\_INELIGIBLE} & provider $\rightarrow$ compliance & provider/compliance \\
Insurance & VAT mismatch & \texttt{REQ\_TAX\_MISMATCH} & compliance $\rightarrow$ legal & compliance/legal \\
Insurance & Data residency & \texttt{REQ\_DATA\_RESIDENCY} & medical $\rightarrow$ fraud & medical/fraud \\
E-commerce & Discount fabrication & \texttt{REQ\_DISCOUNT\_EXCEEDED} & marketing $\rightarrow$ inventory & marketing/inventory \\
E-commerce & Marketplace price drift & G7 keyed price drift & marketplace agents $\rightarrow$ marketing & marketplace/marketing \\
E-commerce & Scope drift & \texttt{REQ\_SCOPE\_EXCEEDED} & marketing $\rightarrow$ inventory & marketing/inventory \\
E-commerce & Revenue inflation & value/tax checks & sales analytics or compliance $\rightarrow$ downstream finance & revenue/tax handoffs \\
B2B & Discount fabrication & value/provenance check & sales $\rightarrow$ onboarding & \texttt{h\_s\_o} \\
B2B & Usage inflation$^{\dagger}$ & usage requirement & support $\rightarrow$ accountancy & \texttt{h\_u\_a} \\
B2B & SLA drift & \texttt{REQ\_TIER\_MISMATCH} & key account $\rightarrow$ support & \texttt{h\_k\_u} \\
Hospital & Patient-ID drift$^{\dagger}$ & G7 / provenance & HPI $\rightarrow$ vitals & HPI/vitals \\
Hospital & Allergy dropped & required allergy checks & pharmacy $\rightarrow$ risk & pharmacy/risk \\
Hospital & Dose inconsistency & reconciliation check & pharmacy $\rightarrow$ risk & pharmacy/risk \\
Enterprise & Cross-handoff inconsistency & G7 entity drift & validator / implementer $\rightarrow$ downstream consumers & multi-hop tracked state \\
\bottomrule
\end{tabularx}
\par\vspace{1mm}{\footnotesize $^{\dagger}$Intended interception point. In the version~1 runs this scenario was not caught by the listed mechanism (v2 audit).}
\end{table}

Two interpretation rules are important. First, the first blocking verdict is not always the mechanism originally associated with the injected profile: an earlier or orthogonal rule can pre-empt a downstream check. Second, a non-firing intended rule does not imply that the workflow was valid; it may have been pre-empted, self-corrected by the model, or misconfigured. These cases are separated explicitly in Appendix B.

\end{document}